\documentclass[12pt]{article} 
\usepackage{amsfonts,epsfig} 
 
\renewcommand{\theequation}{\arabic{section}.\arabic{equation}}

\newcommand{\bra}[1]{\left\langle #1 \right|} 
\newcommand{\ket}[1]{\left| #1 \right\rangle}

\newcommand{\beq}[1]{
\begin{equation}\label{#1}} 
 
\newcommand{\eeq}{\end{equation}}

\newcommand{\beqar}[1]{
\begin{eqnarray}\label{#1}} 
 
\newcommand{\eeqar}{\end{eqnarray}}

\newcommand{\nn}{\nonumber}  
\newcommand{\dd}{{\rm d}}  
\newcommand{\Gl}[1]{Eq.~(\ref{#1})}  
\newcommand{\Ab}[1]{Fig.~\ref{#1}}  
\newcommand{\Ta}[1]{Table~\ref{#1}}  

\newcommand{\D}{{\cal D}}  
  
\newcommand{\V}{{\cal V}}  
\newcommand{\A}{{\cal A}}  
\newcommand{\T}{{\cal T}}  
\newcommand{\up}{{\uparrow}}  
\newcommand{\down}{{\downarrow}}  
\newcommand{\al}{\alpha}  
\newcommand{\be}{\beta}  
\newcommand{\ep}{\varepsilon}  
\newcommand{\ga}{\gamma}  
\newcommand{\de}{\delta}  
  
\newcommand{\la}{\lambda}  
  
\newcommand{\si}{\sigma}  
\newcommand{\ro}{\varrho}

\begin{document}

  
\begin{titlepage}  

\vspace*{-2cm}  
\begin{flushright}  
\begin{tabular}{l}  
hep-ph/0112085  
\end{tabular}  
\end{flushright}  
  
\vskip2.5cm  
  
\begin{center}  
{\Large \bf Light-Cone Sum Rules for the Nucleon Form Factors}  
\vspace{1cm}  
\end{center}  
\centerline{V.M.~Braun$^a$, A.~Lenz$^a$, N.~Mahnke$^a$, E.~Stein$^{a,b}$}  
\vspace{1 cm}  
\centerline{$^a${\em  Institut f{\"u}r Theoretische Physik,  
Universit{\"a}t Regensburg,  
D-93040 Regensburg, Germany}}  
\centerline{$^b${\em  Physics Department,  
Maharishi University of Management,  
NL-6063 NP Vlodrop, Netherlands}}  
\vspace{1cm}  
\bigskip  
\centerline{\large \em \today}  
\bigskip  
\vfill  
\begin{center}  
  {\large\bf Abstract\\[10pt]} \parbox[t]{\textwidth}{  
 We argue that soft non-factorisable terms give a significant contribution  
to the baryon form factors at intermediate  momentum transfers 
and set up a framework for the calculation of such terms 
in the light-cone sum rule approach.
Among them, contributions of three-quark states with different  
helicity structure compared to the leading twist prove to be  
the most important. The leading-order sum rules  
are derived and confronted  with the experimental data. }  
  \vskip1cm  
{\em Submitted to Physical Review D }\\[1cm]  
\end{center}  
  
\vspace*{1cm}  
  
\vspace*{\fill}  
\eject  
\end{titlepage}  
  
  
\section{Introduction}  
\setcounter{equation}{0}  
Electromagnetic form factors of the nucleon present a classical observable 
that characterizes the nucleon's spatial charge and current distributions.  
The earliest investigations of the proton form factor \cite{Hof56}  
established the dominance of the one-photon exchange process  
in the electron-proton scattering. The matrix element of the vector  
current taken between nucleon states is conventionally  written in terms  
of the Dirac and Pauli form factors  
$F_1(Q^2)$ and $F_2(Q^2)$, respectively:  
\beqar{formfactor}  
\bra{P-q} j_\mu^{\rm em}(0) \ket{P} =  
\bar{N}(P-q) \left[ \ga_\mu F_1(Q^2) -  
i \frac{\si_{\mu\nu}q^\nu}{2 M}  F_2(Q^2) \right]N(P)  
\; ,  
\eeqar  
where $P_\mu$ is the nucleon four-momentum in the initial state, $M$ 
is the nucleon mass, $P^2= (P-q)^2=M^2$, 
$q_\mu$ is the (outgoing) photon momentum, $Q^2 = -q^2$,   
and $N(P)$ is the nucleon spinor. 
The values of Dirac and Pauli form factors at $Q^2 = 0$ define the  
electric charge and the anomalous magnetic moment of the nucleon:  
\beqar{zahlen}  
&&F^p_1(0) = 1 \,, \quad F^n_1(0) = 0 \,, \quad  
F^p_2(0) = \kappa_p = 1.79  \,,\quad  
F^n_2(0) = \kappa_n = -1.91 \, .  
\eeqar  
Hereafter, `$p$' and `$n$' stand for the proton and the neutron,  
respectively. From the experimental point of view it is more convenient to  
work with  the electric $G_E(Q^2)$ and magnetic $G_M(Q^2)$ Sachs form  
factors defined as 
\beqar{sachs}  
&& G_{\rm M}(Q^2) = F_1(Q^2) + F_2(Q^2) \,, \qquad  
G_{\rm E}(Q^2) = F_1(Q^2) + \frac{q^2}{4 M^2} F_2(Q^2) \, ,  
\nn \\  
&&G^p_{\rm M}(0) = \mu_p = 2.79 \, ,\qquad\qquad\qquad  
G^n_{\rm M}(0) = \mu_n  = -1.91 \, .  
\eeqar  
In a special frame of reference, the Breit-frame,  
$G_{\rm E}(Q^2)$ corresponds to the distribution of electric charge and 
 $G_{\rm M}(Q^2)$  to the magnetic current distribution.  
In the same frame $G_{\rm M}(Q^2)$  stands for the helicity conserving  
amplitude, while $G_{\rm E}(Q^2)$ corresponds to a helicity-flip.  
In the infinite momentum frame $F_1(Q^2)$ and $F_2(Q^2)$ are helicity  
conserving and helicity violating, respectively. 
  
It is known that the experimental data for $G_M(Q^2)$ at values of  
$Q^2$ up to 5 GeV$^2$  are very well described by the famous dipole formula:  
\beqar{dipole}  
\frac{1}{\mu_p} G_M^{p}(Q^2)  
\sim  
\frac{1}{\mu_n} G_M^{n}(Q^2) \sim  
\frac{1}{(1+Q^2/\mu_0^2)^2}  = G_D(Q^2)\,;
\qquad \mu_0^2 \sim 0.71\,{\rm GeV^2} \; .  
\eeqar  
For the electric form factor a dipole behavior is observed for $Q^2$  
below 1 GeV$^2$. For larger momentum transfers the experimental 
situation was unclear until recently, because of  SLAC \cite{SLAC94} data 
contradicting the older DESY results \cite{DESY73}.  
Both of these measurements were based on the traditional Rosenbluth 
separation of the cross section. Very recently,  
 the Jefferson Lab Hall~A Collaboration determined the ratio  
$G^p_E(Q^2)/G^p_M(Q^2)$ from a simultaneous measurement of longitudinal and  
perpendicular  polarization components of the recoil nucleon 
\cite{JLab00,GEproton}.  
A systematic deviation from the dipole behavior for the electric form factor  
was observed confirming the tendency seen in earlier measurements at DESY.  
 
The theoretical calculation of the form factors from the underlying field  
theory presents a classical problem of the physics of strong interactions. 
{}For very large $Q^2$, the hard gluon exchange contribution proves to  
be dominant. The corresponding formalism was developed in  
\cite{exclusive}, presenting one of the highlights of perturbative QCD. 
This approach introduces the concept of hadron distribution amplitudes  
as fundamental nonperturbative  
functions describing the hadron structure in rare parton configurations  
with a minimum number of Fock constituents (and at small transverse  
separations). The Dirac form factor $F_1(Q^2)$ exhibits the  
leading asymptotic $1/Q^4$ scaling behaviour and can be written in  
a factorized form as a convolution of 
two nucleon distribution amplitudes of leading twist and a calculable  
hard part. On the other hand, the Pauli form factor $F_2(Q^2)$ turns out to 
be  additionally suppressed by an extra power of $1/Q^2$. It is, therefore,  
of higher twist and is not accessible within the standard approach.   
  
In practice the calculations of nucleon form factors for  
realistic values of $Q^2$ using the standard hard-scattering picture  
appear to be not convincing.  
For the asymptotic form of the leading twist three-quark  
distribution amplitude the proton form factor turns out to be zero 
to leading order \cite{earlybaryon}, while the neutron form factor
compared to the data is small  
and of opposite sign. A remedy that has been suggested  
in \cite{Che84} is that nucleon distribution amplitudes at intermediate  
momentum transfers may deviate strongly from their asymptotic shape.  
Chernyak and Zhitnitsky have shown that it is indeed possible to get a 
satisfactory description of the magnetic form factors using very asymmetric 
distribution amplitudes in which a large fraction of the nucleon  
momentum is carried by one valence quark. One drawback of this suggestion  
is, however, that even at large values of  
$Q^2 \sim 10$~GeV$^2$ the ``hard'' amplitude is  
dominated by small gluon virtualities \cite{Isgur}, casting  
doubt on the consistency of the perturbative approach. Attempts have been  
made to increase the region of applicability of perturbative QCD by  
resumming Sudakov-type logarithmic corrections to all orders \cite{LS92}. 
Unfortunately, the Sudakov suppression of large transverse separations is  
most likely not strong enough to suppress nonperturbative effects, 
see e.g. \cite{Kroll} for a detailed discussion. 
 
Another point of view that is becoming increasingly popular  
in recent years is that the onset of the 
perturbative QCD regime in exclusive reactions is postponed until  
very large momentum transfers and nonperturbative   
 so-called ``soft'' or ``end-point'' contributions to exclusive reactions  
play a dominant role at present energies. In particular, it proves to  
be possible to get a good description of the existing data by the ``soft'' 
contribution alone, modeled by an overlap of nonperturbative  
wave functions, see \cite{softnucleon}. A weak point of this approach  
is a possibility of double counting, with hard rescattering contributions  
``hidden'' in model-dependent hadron wave functions. 
 
In this paper we develop an 
approach to the calculation of baryon form factors based on light-cone  
sum rules (LCSR) \cite{LCSR}. Although the LCSR predictions do involve a  
certain model dependence and the leading-order sum rules may  
be not very accurate, this technique offers an important advantage  
of being fully consistent with QCD perturbation theory. LCSRs reveal that 
the distinction between ``hard'' and ``soft'' contributions appears to 
be scale- and scheme-dependent \cite{BKM00}. It was demonstrated for  
the case of the pion \cite{BKM00} that the contribution of hard  
rescattering is correctly reproduced in the LCSR   
approach as a part of the $O(\alpha_s)$ correction.   
In recent years there have been numerous applications of LCSRs  
to mesons, see \cite{LCSRreview} for a review. Baryon form factors, 
however, were never considered. 
One reason for this is that the LCSR calculations 
require certain knowledge about the distribution amplitudes of higher  
twists and for baryons these were not available until recently \cite{BFMS}. 
Another reason is that, as we will see, the LCSR formalism for baryons  
appears to be considerably more cumbersome. 
 
Apart from resolving several technical issues, our main finding in this work  
is that the soft contribution to the nucleon form factors is   
dominated by valence quark configurations with different helicity  
structure compared to the leading-twist amplitude. Large contributions of the  
distribution amplitudes with ``wrong'' helicity are important for the  
electric form factor and in a more general context can explain why  
helicity selection rules in perturbative QCD appear to be badly  
broken in hard exclusive processes at present energies.      
The sum rules in the present paper are derived to leading order 
in the QCD coupling. Comparing the results with the available data we conclude
that nucleon distribution amplitudes that deviate significantly  
from their asymptotic shape are disfavored. With the asymptotic distribution 
amplitudes, the accuracy of the sum rules proves to be of order 50\%  
in the range  $1<Q^2<10$~GeV$^2$ both for the proton and the neutron.  
We believe that the accuracy can be improved significantly by the  
calculation of $O(\alpha_s)$ corrections to the sum rules and especially  
if lattice data on the moments of higher-twist distribution amplitudes  
become available.     
  
The presentation  is organized as follows.
In Section 2 we introduce the necessary notation and explain basic 
ideas and techniques of the LCSR approach on the example of the 
leading-twist contribution to the sum rule.
Section 3 contains the derivation of sum rules including higher 
twist corrections, which is our main result.
The numerical analysis of the LCSRs is carried out in Section~4,
together with a summary and discussion.
The paper has two Appendices devoted to technical aspects of the 
calculation: In Appendix A we collect the necessary expressions 
for the conformal expansions of nucleon distribution amplitudes
and in Appendix B a derivation of nucleon mass corrections to
the sum rules is given.


\section{Getting Started: Leading Twist}  
\setcounter{equation}{0}  
  
The method of light-cone sum rules \cite{LCSR} combines the standard  
technique of QCD sum rules \cite{SVZ} with the specific light-cone  
kinematics of hard exclusive processes. As the most apparent distinction, 
the short-distance Wilson operator product expansion in contributions  
of vacuum condensates of increasing dimension is replaced by the  
light-cone expansion in terms of distribution amplitudes  of increasing twist.  
Our calculation is similar to the calculation of the pion  
form factor in Ref.~\cite{BKM00}.  
  
Throughout this work we consider the following correlation function  
\beqar{correlator}  
T_\nu(P,q) = i \int \dd^4 x \, e^{i q \cdot x}  
\bra{0} T\left\{\eta(0) j_\nu^{\rm em} (x)\right\} \ket{P}  
\eeqar  
which includes the electromagnetic current  
\beqar{current}  
j_\nu^{\rm em} = e_u \bar{u} \ga_\nu u + e_d \bar{d} \ga_\nu d  
\eeqar  
and an interpolating nucleon (proton) field \cite{Che84}  
\beqar{local-operators}  
\eta_{\rm CZ}(0) &=& \ep^{ijk} \left[u^i(0) C \!\not\!{z} u^j(0)\right] \,  
\ga_5 \!\not\!{z} d^k(0) \,, 
\nonumber\\ 
 \bra{0} \eta_{\rm CZ}  \ket{P} & = & f_{\rm N}\,  
(P \cdot z) \!\not\!{z} N(P)\,.  
\eeqar  
Here $z$ is a light-cone vector, $z^2 = 0$,  
and the coupling $f_N$ determines the normalization of the leading twist 
proton distribution amplitude \cite{earlybaryon}.  
This choice is convenient for our purposes as it leads to the same hierarchy 
of the contributions of different twists as in the perturbative approach 
\cite{exclusive}, see below. 
From the definition in \Gl{formfactor} the contribution of the nucleon  
intermediate state in the correlation function  
\Gl{correlator} is readily derived to be   
\beqar{dispersion}  
z^\nu T_\nu(P,q) &=& \frac{f_N}{M^2- {P'}^2} (P'\cdot z) 
\bigg\{\bigg[ 2 F_1(Q^2) \left(P'\cdot z\right) - F_2(Q^2) (q\cdot z)\bigg]  
\!\not\!{z}\,  
\nn \\ && \hspace*{3.2cm} 
{}+ F_2(Q^2)\,\left[ (P'\cdot z) + \frac12 (q\cdot z)\right]  
\frac{\!\not\!{z} \!\not\!{q}}{M}  
\bigg\} N(P) + \ldots  
\eeqar  
where  
\beqar{P'} 
 P' &=& P-q 
\eeqar 
and the dots stand for higher resonances and the continuum.  
In order to simplify the Lorentz structure we have contracted the correlation  
function with $z^\nu$ to get rid of contributions $\sim z_\nu$ that give  
subdominant contributions on the light-cone.

\begin{figure}[t]  
\centerline{\epsfig{file=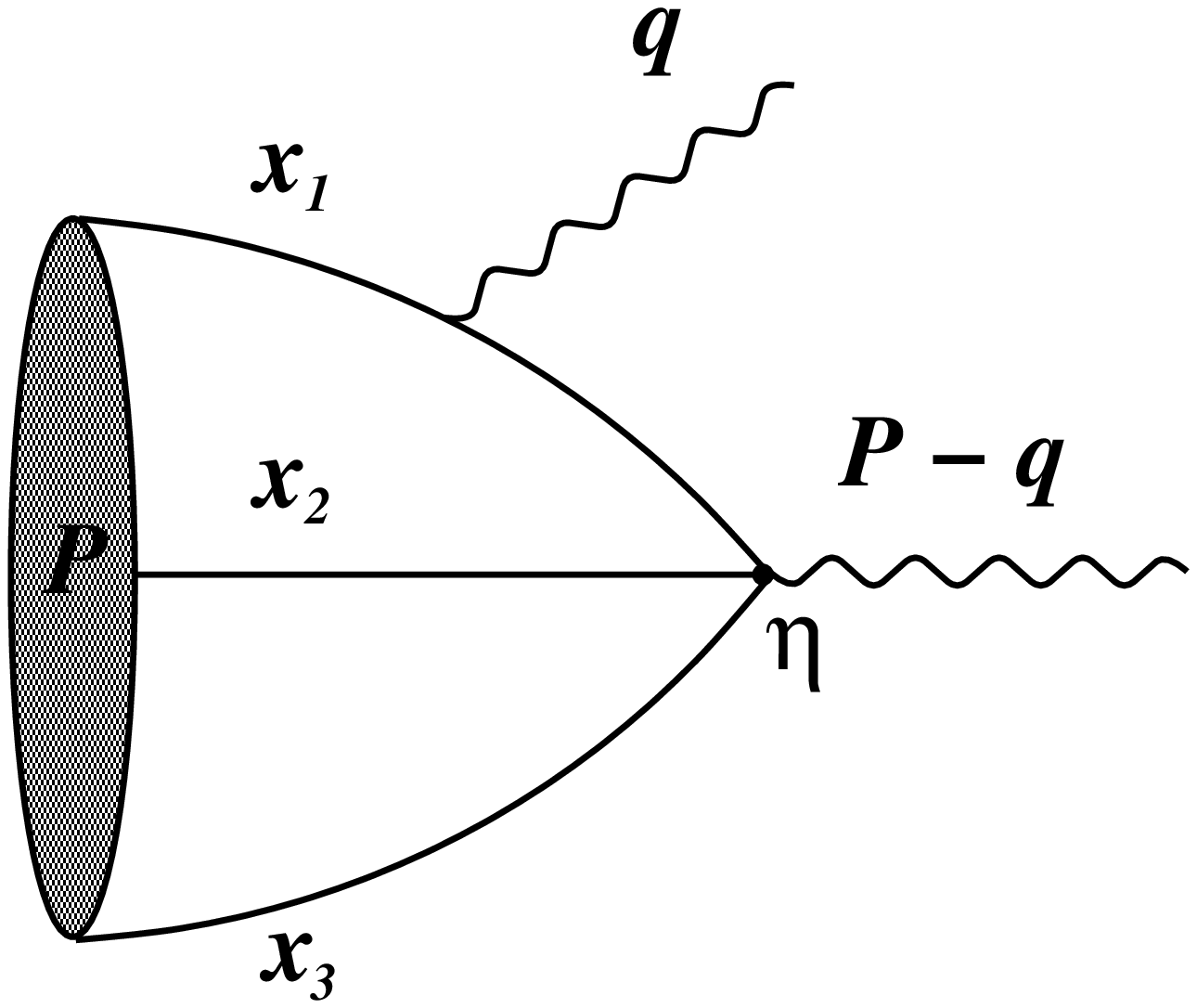,width=6.0cm}}   
\caption[]{\small \sf The tree-level contribution to the correlation function  
\Gl{correlator}.}  
\label{figure1}  
\end{figure}  
On the other hand, at large Euclidean momenta  
${P'}^2$ and $q^2 = - Q^2$ the correlation function can  
be calculated in perturbation theory. The leading order  
contribution is obtained from the diagram shown in \Ab{figure1}.  
A simple calculation yields  
\beqar{calc1}  
z_\nu T^\nu(P,q) &=&  
\frac{1}{2\pi^2}\int \dd^4 x\, \frac{e^{iqx}}{x^4}  
\left(C \!\not\!{z}\right)_{\al\be} (z\cdot x)  
\left(\ga_5  \!\not\!{z}\right)_\ga  
\nn \\  
&& \times  
\bigg[2 e_{\rm d}\bra{0}\ep^{ijk} u_\al^i(0) u_\be^j(0) d_\ga^k(x)  \ket{P} +  
4e_{\rm u}\bra{0}\ep^{ijk}u_\al^i(0)u_\be^j(x)d_\ga^k(0)\ket{P} \bigg]  
\,,  
\eeqar  
where $\alpha,\beta,\gamma$ are quark spinor indices.  
In the light-cone limit $x^2\to 0$ the remaining 
three-quark operator sandwiched between the proton state and the vacuum  
can be written in terms of the leading twist nucleon distribution  
amplitudes \cite{earlybaryon,Che84,BFMS} 
\beqar{ltwistwf}  
 4 \bra{0} \ep^{ijk} u_\al^i(a_1 z) u_\be^j(a_2 z) d_\ga^k(a_3 z)  
          \ket{P}  
&=&  
\V_1  \left(\!\not\!{P}C \right)_{\al \be} \left(\ga_5 N\right)_\ga +  
\A_1  \left(\!\not\!{P}\ga_5 C \right)_{\al \be} N_\ga  
\nn \\  
&& {}+  
\T_1 \left(P^\nu i \si_{\mu\nu} C\right)_{\al \be}  
\left(\ga^\mu\ga_5 N\right)_\ga  
\,.  
\eeqar  
Each distribution amplitude $\V_1, \A_1$ and $\T_1$  can be represented  
as  
\beqar{representation}  
F(a_k p\cdot z) = \int  {\cal D} x \, e^{-i pz \sum_j x_j a_j} F(x_i)\,.  
\eeqar  
The integration runs over the longitudinal momentum fractions  
$x_1, x_2, x_3$ carried by the quarks  
inside the nucleon with $\sum_i x_i = 1$ and the integration measure 
is defined as  
\beqar{measure}  
\int {\cal D} x  = \int_0^1\dd x_1\dd x_2 \dd x_3 \,\de(x_1+x_2 +x_3 -1) \,.  
\eeqar  
The normalization is fixed by  
\beqar{normierung}  
\int {\cal D} x \, \V_1(x_1,x_2,x_3) = f_N \,,  
\eeqar  
cf. \Gl{local-operators}. 
With these definitions, we find for the contribution in \Gl{calc1}  
\beqar{ltwist}  
z_\nu T^\nu = - \left[e_d \int \D x \frac{x_3 \V_1(x_i)}{(q- x_3 P)^2} +  
2 e_u \int \D x \frac{x_2 \V_1(x_i)}{(q- x_2 P)^2}\right]  
2 (P\cdot z)^2 \!\not\!{z}\, N(P)  + \ldots  
\eeqar  
where the ellipses stand for contributions that are nonleading in the  
infinite momentum frame kinematics $P\to\infty, q\sim const, z \sim 1/P$. 
Note that in this case only $\V_1$ contributes and there is no 
Lorentz structure  
$\sim (P\cdot z)^2 \!\not\!{z} \!\not\!{q}\,$ that would give rise to  
the Pauli form factor $F_2$.  
 
The common idea of QCD sum rules is  to match the dispersion representation  
in \Gl{dispersion} with the QCD calculation at certain ``not so large''  
Euclidian values of the momentum ${P'}^2$ flowing through the nucleon  
interpolation current.  
To this end we can rewrite the perturbative  
result in \Gl{ltwist} in the form of a dispersion relation with a certain  
spectral density  
\beqar{rho}  
z^\nu T_\nu(P,q) = \int\limits_0^{s_0} \dd s \, 
\frac{\ro(s,Q^2)}{s -{P'}^2} \; 2 (P\cdot z)^2 \!\not\!{z}\, N(P)  
+ \ldots      \,.  
\eeqar  
Restricting the region of integration to the mass region below the Roper  
resonance, $s_0 \sim (1.5  \,{\rm GeV})^2$,  
one eliminates contributions other than the nucleon \cite{SVZ}. 
If terms of ${\cal O}(M^2/Q^2)$ are neglected 
--- which is consistent with twist-3 accuracy --- 
such a representation is obtained easily by the  
substitution $s = (1- x_3) Q^2/ x_3$ or $s = (1- x_2) Q^2/x_2$ for the  
contribution of the $d$-quark and the $u$-quark, respectively.   
The upper bound in the dispersion integral then translates into a lower bound  
in the integral over the corresponding momentum fraction:  
$ x_2 > Q^2/(s_0 +Q^2)$ or $x_3 > Q^2/(s_0 +Q^2)$.  
{}Finally  we follow  
the usual QCD sum rules procedure to use a Borel transformation  
to convert the power suppression of higher mass contributions  
into an exponential suppression  
\beqar{borel}  
\frac{1}{-(q-x  P)^2} &=& \frac{1}{x (s  - {P'}^2)}  
\to \frac{1}{x} \exp \left\{- \frac{s}{M_B^2}\right\}  
\, .  
\eeqar  
As the result, a new variable --- the Borel parameter $M_B$ --- enters  
instead of ${P'}^2$.  
Equating the Borel transformed versions of \Gl{dispersion}  
and \Gl{ltwist} we finally  arrive at the sum rules 
\beqar{sumlt}  
F^{\rm tw-3}_1(Q^2) &=& \frac{1}{f_N}  
\left[e_d \int \D x\,  \V_1(x_i)  
\exp{  
\displaystyle  
\left(- \frac{ \bar x_3 Q^2 - x_3^2 M^2}{x_3 M_B^2} \right)}  
\Theta\left(x_3 - \frac{Q^2}{Q^2 + s_0}\right)  
\nn \right.\\ && \left.  
{}+2 e_u \int \D x\, \V_1(x_i)  
\exp{  
\displaystyle  
\left(- \frac{ \bar x_2 Q^2 - x_2^2 M^2}{x_2 M_B^2} \right)}  
\Theta\left(x_2 - \frac{Q^2}{Q^2 + s_0}\right)  
\right]  
\,, 
\nn \\ 
F^{\rm tw-3}_2(Q^2) &=& 0\,. 
\eeqar  
where the superscript `tw-3' reminds of twist-three accuracy%
\footnote{Throughout this paper we give explicit expressions for the
proton. Neutron form factors are obtained by the obvious substitution 
$e_u\leftrightarrow e_d$.}.
 
The $\Theta$-functions in \Gl{sumlt} originate from the restriction  
of the spectral density to the duality region $s<s_0$ and confine  
the integration region to values of $x_3 \to 1$ and $x_2 \to 1$ at  
$Q^2 \to \infty$ for the contributions of $d$ and $u$ quarks,
respectively.  
It follows that in our approximation the form factor is dominated by parton  
configurations where the scattered quark carries almost all the momentum  
of the nucleon. This is precisely the soft, or Feynman mechanism,  
and the standard wisdom tells that this contribution  
has to be subleading at very large momentum transfers. Indeed,  
using the asymptotic wave function $\V_1(x_i) = 120 f_N x_1 x_2 x_3$  
and expanding in $1/Q^2$ we find that the sum rule result in the limit 
$Q^2\to\infty$ behaves as 
\beqar{ltexpand}  
F^{\rm tw-3}_1(Q^2) &=& 20 \left(e_d + 2 e_u\right) \frac{1}{Q^8}  
\int_0^{s_0} \dd s  \, s^3 e^{ -\frac{s - M^2}{M_B^2}}  
\,,  
\eeqar  
i.e. it is suppressed by {\em two} additional powers of $1/Q^2$ compared  
with the expected asymptotic behavior.
This observation is in agreement with an analysis of the soft 
contribution in the framework of the Drell-Yan description of $F_1$ \cite{DY}:
The soft overlap contribution of two leading twist light-cone wave 
functions computed in \cite{DFJK} asymptotically vanishes as $1/Q^8$.
We will find, however, that this strong suppression does not hold for  
contributions of wave functions with different helicity structure. 
\begin{figure}
\centerline{  \epsfig{file=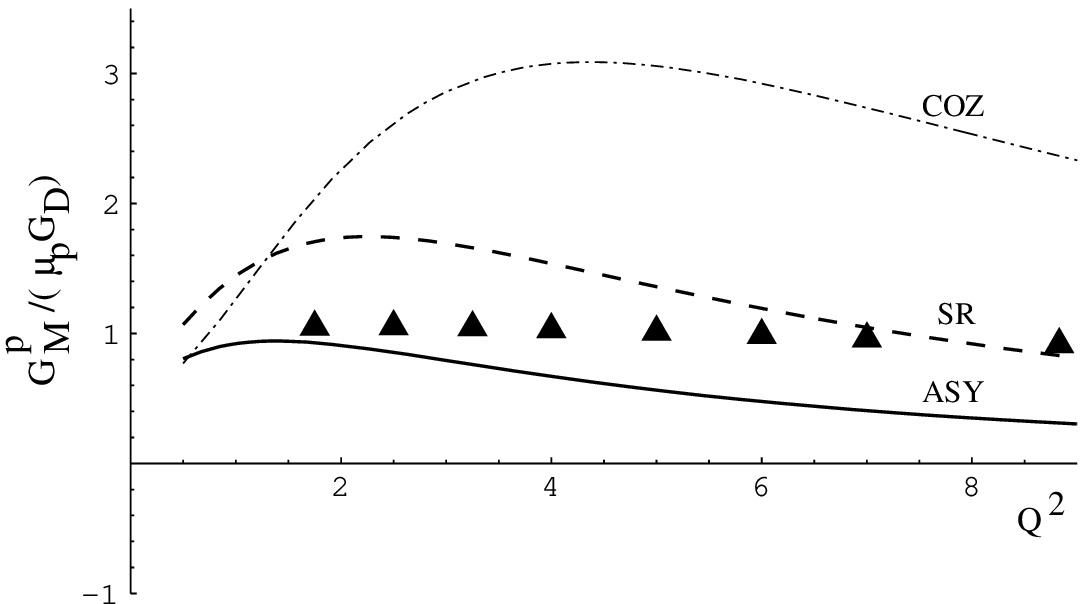,width=7.5cm}\qquad
              \epsfig{file=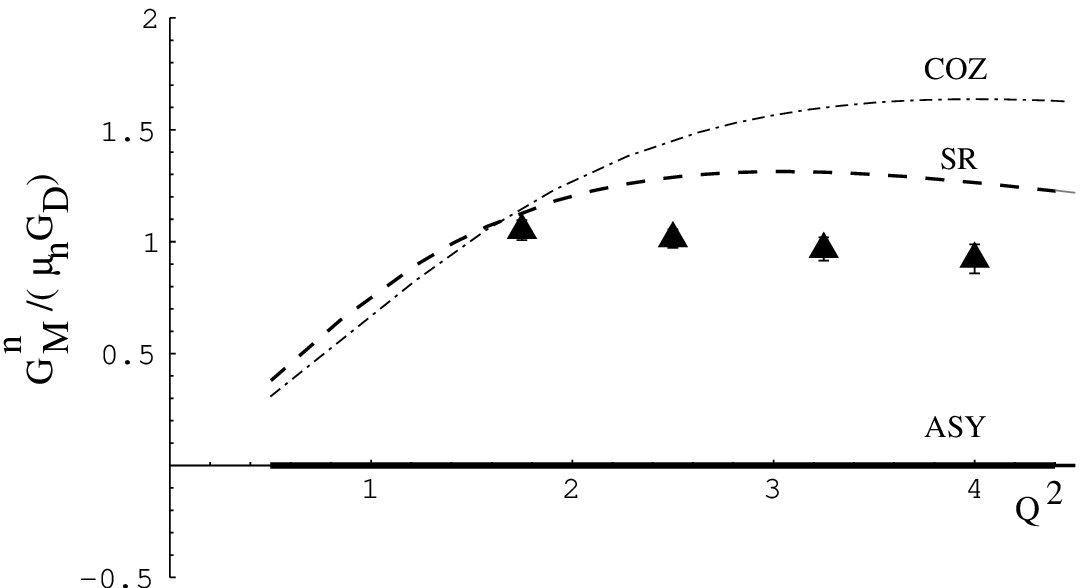,width=7.5cm}  }
\caption[]{\small\sf Twist-3 approximation to the soft contribution 
to the magnetic form factor of the
proton (left)  and the neutron (right), see \Gl{sumlt}. 
Plotted is the ratio 
$G_M^{\rm p/n}/(\mu_{\rm p/n} G_D)$ where $G_D$ is the dipole fit 
\Gl{dipole}. The three curves correspond to different models of 
the leading twist distribution amplitudes: The full line  corresponds to 
${\cal V}_1^{\rm asy}$, the dashed line to ${\cal V}_1$, and 
the dashed-dotted line to ${\cal V}_1^{\rm COZ}$, respectively,
see \Gl{da-twist-3}. Sum rule parameters are taken to be 
$M_B^2 = 2\, \rm{GeV^2}$ and $s_0 = (1.5\, {\rm GeV})^2$. The data points are
taken from \cite{protonGM,neutron}.}  
\label{figure2}  
\end{figure}

It is  well known that the form factor is strongly sensitive 
to the shape of the nucleon distribution amplitude. The general QCD 
description of distribution amplitudes is 
based on the conformal expansion \cite{exclusive,earlybaryon,BDKM}.
There exist several concrete models that
take into account the first few conformal partial waves. 
In \Ab{figure2} we show  the twist-3 LCSR prediction (\Gl{sumlt}) for the
magnetic form factor $G^{p,n}_M$ (to this accuracy $G_M =F_1$) 
normalized to the phenomenological dipole parametrization \Gl{dipole},
for three different choices of the distribution amplitude
\beqar{da-twist-3}
{\cal V}_1^{\rm asy}(x_i, \mu\simeq 1 {\rm GeV}) 
&=& 120 x_1 x_2 x_3 f_N \, ,\nn \\
{\cal V}_1 (x_i, \mu\simeq 1 {\rm GeV}) 
&=& 120 x_1 x_2 x_3 f_N \left[1 + \tilde{\phi}_3^+(\mu)(1-x_3)
\right]\, , \nn \\
{\cal V}_1^{\rm COZ} (x_i, \mu\simeq 1 {\rm GeV}) 
&=& 120 x_1 x_2 x_3 f_N \bigg[1 + \tilde{\phi}_3^+(\mu)(1-x_3)
\nn \\
&&{}+\tilde\phi_3^{d1}\big[3-21 x_3+28 x_3^2\big] 
    +\tilde\phi_3^{d2}\big[5(x_1^2+x_2^2)-3(1-x_3)^2)\big]\bigg] 
\,,
\eeqar
where 
\beqar{twist3par}
f_N(\mu =1 {\rm GeV}) &=& (5.3 \pm 0.5) \times 10^{-3} {\rm GeV^2} \,,
\nn \\
\tilde{\phi}_3^+(\mu = 1 {\rm GeV}) &=& 1.1\pm 0.3\,,
\nn \\
 \tilde\phi_3^{d1}(\mu = 1 {\rm GeV}) &=& 0.615\,,
\nn \\
 \tilde\phi_3^{d2}(\mu = 1 {\rm GeV}) &=& 3.68\,.
\eeqar
The first expression in \Gl{da-twist-3}
defines the asymptotic distribution amplitude in the limit 
$Q^2\to\infty$ and corresponds to the contribution of the lowest conformal 
spin (``S-wave'') \cite{earlybaryon}. The second and the third expressions
correspond to taking into account the next-to-lowest (``P-wave'')
and also the next-to-next-lowest (``D-wave'') conformal spin, respectively%
\footnote{We have rewritten the original expressions given in \cite{Che84} 
using the conformal basis defined in the work \cite{BDKM}. 
In this basis, all terms are mutually orthogonal with respect to the
 conformal weight function ${\cal V}_1^{\rm asy}(x_i)$.}.
The parameters in \Gl{twist3par} are taken from the last reference in
\cite{Che84}. They were obtained using QCD sum rules for the lowest moments
of the distribution amplitudes and their accuracy is subject of debate. 
Note that the dependence on $f_N$ in fact cancels out of the sum rule 
in \Gl{sumlt}. The numerical impact of the scale dependence of the 
remaining parameters is negligible compared to the intrinsic uncertainty of 
their estimates, so that we do not take this effect into account.

{}From \Ab{figure2} it is seen that the results vary by factor 2--3
depending on the choice of the distribution amplitude. The ``P-wave''
approximation in the second line in \Gl{da-twist-3} seems to fit best
while for the asymptotic distribution amplitude the neutron form factor 
turns out to be zero. One has to keep in mind, however, 
that in this calculation $F_2(Q^2)$ is zero both for proton and neutron
and for arbitrary choice of the distribution amplitude. Hence we have, e.g.
$G_M(Q^2) = G_E(Q^2)$ identically, which is clearly not a satisfactory 
approximation in a few GeV region. We conclude that the leading twist
approximation to the LCSRs is not sufficient for the quantitative
analysis and in the next section we proceed to the complete treatment, 
including higher-twist effects.

\section{Beyond the Leading Twist}  
\setcounter{equation}{0}  
  
Apart from radiative corrections, the correlation function in \Gl{correlator}
receives further contributions of higher-twist distribution amplitudes. 
These contributions arise in two ways. First, one has to remember that 
the hard quark propagator $\sim\, \not\! x\,/\,x^4$ in \Gl{calc1} receives 
corrections in the background color field \cite{BB89}, which  
are proportional to the gluon field strength tensor and give rise to 
four-particle (and five-particle) nucleon distribution amplitudes 
corresponding to parton states with a gluon or quark-antiquark
pair in addition to the three valence quarks. Such corrections are usually 
not expected to play any significant r{\^o}le (see e.g. \cite{DFJK}), 
and in this work 
we do not take them into account. Second, one has to improve on the 
treatment of the matrix elements of the three-quark operators in \Gl{calc1}
taking into account contributions of other Lorentz structures and less 
singular contributions on the light-cone, 
beyond the leading-twist approximation in \Gl{ltwistwf}. 

We only need to retain vector-like Dirac structures since it is easy to see  
(cf. \Gl{calc1}) that all others do not contribute to the sum rule we are 
considering. To this accuracy, the complete decomposition of the 
three-quark operator reads \cite{BFMS}
\beqar{zerl}  
\lefteqn{ 4 \bra{0} \ep^{ijk} u_\al^i(a_1 x) u_\be^j(a_2 x) d_\ga^k(a_3 x)  
          \ket{P} =}  
 \\  
&=&  
\left({\cal V}_1 + \frac{x^2M^2}{4} \V_1^M \right) \left(\!\not\!{P}C \right)_{\al \be} \left(\ga_5 N\right)_\ga +  
\V_2 M \left(\!\not\!{P} C \right)_{\al \be} \left(\!\not\!{x} \ga_5 N\right)_\ga  +  
\V_3 M  \left(\ga_\mu C \right)_{\al \be}\left(\ga^{\mu} \ga_5 N\right)_\ga  
\nn \\  
&& +  
\V_4 M^2 \left(\!\not\!{x}C \right)_{\al \be} \left(\ga_5 N\right)_\ga +  
\V_5 M^2 \left(\ga_\mu C \right)_{\al \be} \left(i \si^{\mu\nu} x_\nu \ga_5  
N\right)_\ga  
+ \V_6 M^3 \left(\!\not\!{x} C \right)_{\al \be} \left(\!\not\!{x} \ga_5 N\right  
)_\ga\, .\nn  
\eeqar   
The expansion in \Gl{zerl} has to be viewed as an operator product expansion 
to the leading order in the strong coupling. Each of the 
functions $\V_i$ depends on the deviation from the light-cone at most 
logarithmically, and we have also included the $O(x^2)$ correction to the 
leading-twist-3 structure, denoted by $\V^M_1$. 
The invariant functions $\V_1(x_i),\ldots,\V_6(x_i)$ 
can be expressed readily in terms of higher-twist nucleon distribution 
amplitudes  $\V_1(x_i),\ldots,\V_6(x_i)$ introduced in Ref.~\cite{BFMS}:
\beqar{opev}  
\renewcommand{\arraystretch}{1.7}  
\begin{array}{lll}  
 \V_1 = V_1\,, &~~& 2 p\cdot x \V_2 = V_1 - V_2 - V_3\,, \\  
 2\, \V_3 = V_3\,, &~~& 4 p\cdot x \V_4 = - 2 V_1 + V_3 + V_4  + 2 V_5\,, \\  
4\, p\cdot x\, \V_5 = V_4 - V_3\,, &~~&  
(2 p\cdot x )^2  \V_6 = - V_1 + V_2 +  V_3 +  V_4 + V_5 - V_6\,.
\end{array}
\renewcommand{\arraystretch}{1.0}  
\eeqar  
In difference to the ``calligraphic'' functions $\V_1(x_i),\ldots,\V_6(x_i)$
each of the distribution amplitudes $V_1(x_i),\ldots,V_6(x_i)$ 
has definite twist, see \Ta{tabelle1}, and corresponds to 
\begin{table}
\renewcommand{\arraystretch}{1.3}
\begin{center}
\begin{tabular}{l|l|l|l}
twist-3  &  twist-4  & twist-5       & twist-6  \\ \hline
$V_1$    & $V_2\;,\;V_3$ & $V_4\;,\;V_5 $& $V_6$ \\ \hline
\end{tabular}
\end{center}
\caption[]{\sf Twist classification of the distribution amplitudes
in \Gl{opev}.}
\label{tabelle1}
\renewcommand{\arraystretch}{1.0}
\end{table}
the matrix element of a (renormalized) three-quark operator with 
exactly light-like separations $x^2\to 0$, see Table~2 and Appendix C in 
Ref.~\cite{BFMS} for the details. The higher-twist distribution 
amplitudes $V_2(x_i),\ldots,V_6(x_i)$ correspond to ``wrong'' components 
of the quark spinors and have different helicity structure compared to 
the leading twist amplitude. For
baryons these ``bad'' components cannot all be traded for gluons as 
in the case of mesons \cite{badmesons}. 
They are not all independent, but related to
each other by the exact QCD equations of motion. As the result,  to 
the leading conformal spin accuracy the five functions 
$V_2(x_i),\ldots,V_6(x_i)$ involve only one single nonperturbative 
higher twist parameter. 
In the calculations presented
below we use the conformal expansions of higher twist distribution amplitudes
to the next-to-leading order (include ``P-wave'').
This accuracy is consistent with neglecting multiparton components 
with extra gluons (quark-antiquark pairs) that are of yet higher spin. 
Explicit expressions for the distribution 
amplitudes are collected in Appendix~A. 

The invariant function $\V_1^M$ corresponding to the  corrections of  
order ${\cal O} (x^2)$ to the leading -twist Lorentz structure in 
\Gl{zerl} is twist-5 and in general is a complicated function 
of distribution amplitudes. In the present context there is an important 
simplification that two of the quarks in \Gl{calc1} always appear to be 
at the same space-time point, cf. \Ab{figure1}. 
In this specific configuration, 
$\V_1^M$ can be obtained using the technique of Ref.~\cite{BB89}
that has been designed for the studies of mesonic operators.
The corresponding calculation is carried out in Appendix~B and the 
expressions for  $\V_1^M$ in the above mentioned limits
present one of the new results of this paper.   
Substituting the full expansion \Gl{zerl} in \Gl{calc1} we obtain the  
result:
\beqar{fullresult}  
z_\nu T^\nu &=& 2 (P\cdot z)^2 \!\not\!{z} N(P)   
\Bigg\{e_d \bigg[ - \int \D x \frac{x_3 V_1(x_i)}{(q- x_3 P)^2} 
\\
&&\hspace*{0.5cm}{}+  
M^2 \int_0^1 \dd x_3 \, x_3^2  
\frac{\left(2 \tilde V_1 - \tilde V_2 - \tilde V_3 -  \tilde V_4 -  
\tilde V_5 
 - \frac{2}{x_3} \V_1^{M(d)} 
\right)(x_3)  
}{(q- x_3 P)^4}  
\nn  \\ && \hspace*{0.5cm} 
{}+  
2 M^4\! \int_0^1\!\! \dd x_3 \, x_3^3  
\frac{  
\left(\tilde{\tilde V}_1 - \tilde{\tilde V}_2 - \tilde{\tilde V}_3 -  
\tilde{\tilde V}_4 - \tilde{\tilde V}_5 + \tilde{\tilde V}_6\right)(x_3)  
}{(q- x_3 P)^6  
} \bigg] + 2 e_u \bigg[ x_3 \leftrightarrow x_2\bigg]  
\Bigg\}  
\nn \\ &&  
+ \frac{(P\cdot z)^2}{M} \!\not\!{z} \!\not\!{q} N(P) 
\Bigg\{e_d \bigg[ M^2 \int_0^1 \dd x_3 \,2  x_3  
\frac{\left(\tilde V_2 + \tilde V_3 -  \tilde V_1 \right)(x_3)  
}{(q- x_3 P)^4}  
\nn  \\ && \hspace*{0.5cm}  
{}-  
4 M^4 \!\int_0^1\!\!\! \dd x_3 \, x_3^2  
\frac{  
\left(\tilde{\tilde V}_1 \!-\! \tilde{\tilde V}_2 \!-\! \tilde{\tilde V}_3 
\!-\!  
\tilde{\tilde V}_4 \!-\! \tilde{\tilde V}_5 + \tilde{\tilde V}_6\right)(x_3)  
}{(q- x_3 P)^6  
} \bigg]  
+ 2 e_u \bigg[ x_3 \leftrightarrow x_2\bigg]  
\Bigg\}   + \ldots \nn\,.   
\eeqar  
We see that at this stage a contribution to $F_2(Q^2)$ arises.  
The functions $\V_1^{M(d)}$ and $\V_1^{M(u)}$ (the second one appears 
 in the u-quark contribution) are given in \Gl{VM} and \Gl{VM2}, while 
the distribution amplitudes with a `tilde' are defined as  
\beqar{tildefunction}  
\tilde{V}(x_3) &=& \int_1^{x_3} \dd x_3' \int_0^{1-x_3'} \dd x_1  
V(x_1,1-x_1-x_3',x_3') \,,  
\nn \\  
\tilde{V}(x_2) &=& \int_1^{x_2} \dd x_2' \int_0^{1-x_2'} \dd x_1  
V(x_1,x_2',1-x_1-x_2')  \,,  
\nn \\  
\tilde{\tilde{V}}(x_3) &=& \int_1^{x_3} \dd x_3'\int_1^{x_3'}  
\dd x_3'' \int_0^{1-x_3''} \dd x_1  
V(x_1,1-x_1-x_3'',x_3'')  \,,  
\nn \\  
\tilde{\tilde{V}}(x_2) &=& \int_1^{x_2} \dd x_2'\int_1^{x_2'}  
\dd x_2'' \int_0^{1-x_2''} \dd x_1  
V(x_1,x_2'',1-x_1-x_2'') \,,  
\eeqar  
and result from partial integration in $x_2$ or $x_3$, respectively:
The integration by parts in $x_2$ or $x_3$ is done in order  
to eliminate the $1/ P\cdot x$ factors that appear in \Gl{zerl} when 
the distribution amplitudes \Gl{opev} are inserted. After this, 
the $\int d^4 x$ integration becomes trivial. 
The surface terms sum up to zero.  

The Borel transformation and the continuum subtraction  
are performed by using the following substitution rules:  
\beqar{borel2}  
\int \dd x \frac{\varrho(x) }{(q-x  P)^4} &=&  
\int_0^1 \frac{\dd x }{x^2} \frac{\varrho(x)}{(s - {P'}^2)^2}  
\nn \\  &\to&  
\frac{1}{M_B^2} \int_{x_0}^1 \frac{\dd x}{x^2} \varrho(x)  
\exp{
\left( - \frac{\bar x Q^2}{x M_B^2}  - \frac{\bar x M^2}{M_B^2}\right)}    
+  
\frac{\varrho(x_0)\,e^{-s_0 /M_B^2} }{Q^2 + x_0^2 M^2} \, ,    
\eeqar  
\beqar{borel4}  
\int \dd x \frac{\varrho(x) }{(q-x  P)^6} &=&  
- \int_0^1 \frac{\dd x }{x^3} \frac{\varrho(x)}{(s - {P'}^2 )^3}  
\nn \\  &\to & -  
\frac{1}{2 M_B^4} \int_{x_0}^1 \frac{\dd x}{x^3} \varrho(x)  
\exp{  
\left( - \frac{\bar x Q^2}{x M_B^2} - \frac{\bar x M^2}{M_B^2}  
\right)}  
- \frac12  
\frac{\varrho(x_0)\,e^{-s_0 /M_B^2} }{x_0\left(Q^2 + x_0^2 M^2\right) M_B^2}  
\nn \\  &&  
+ \frac12  \frac{x_0^2}{Q^2 + x_0^2 M^2} \left[\frac{d}{dx_0}  
\frac{\varrho(x_0)}{x_0\left(Q^2 + x_0^2 M^2\right)} \right]  
\,e^{-s_0 /M_B^2} \, \nn \\  
\eeqar  
where in difference to \Gl{sumlt} we have to keep the nucleon mass 
in the kinematical relation $s = \frac{1-x}{x} Q^2  + (1-x) M^2$ and 
$x_0$ is the solution of the corresponding quadratic equation for $s = s_0$:
\beqar{x0}
   x_0 &=&\bigg[ \sqrt{(Q^2+s_0-M^2)^2+ 4 M^2 Q^2}-(Q^2+s_0-M^2)\bigg]
   /(2M^2)\,.
\eeqar  
The contributions $\sim e^{-s_0 /M_B^2}$ in \Gl{borel2} and \Gl{borel4}
correspond to the ``surface terms'' arising from  successive  
partial integrations to reduce the power in the denominators  
$(q - x P)^{2n} = (s  - {P'}^2 )^{2n} (-x)^{2n}$  
with $n > 1$ to the usual dispersion representation with the denominator  
$\sim (s - {P'}^2 )$. Without continuum subtraction, i.e. in the 
limit  $s_0 \to \infty$ these terms vanish.  

Collecting everything together, we get the final expressions:  
 \beqar{F1}  
 F^{\rm p}_1(Q^2) &=&  
 \frac{e_d}{f_N}\Bigg\{ \int_{x_3^0}^1 \dd x_3  
 \exp{\left(- \frac{ \bar x_3 Q^2}{x_3 M_B^2} + \frac{x_3 M^2}{M_B^2}\right)}  
 \Bigg[  
 \left(\int_0^{1\!-\!x_3}\!\! \dd x_1 V_1(x_1,1\!-\!x_1\!-\!x_3,x_3)\right)  
 \nn \\ && \hspace*{2cm}
 {}+ \frac{M^2}{M_B^2}  
 \left(2 \tilde V_1 \!-\! \tilde V_2 \!-\! \tilde V_3 \!-\!  \tilde V_4 \!-\!  
 \tilde V_5 - \frac{2}{x_3}\V_1^{M(d)} \right)(x_3)  
 \nn \\ && \hspace*{2cm}  
 + \frac{M^4}{M_B^4}  
 \left(\tilde{\tilde V}_1 \!-\! \tilde{\tilde V}_2 \!-\! \tilde{\tilde V}_3 \!-\!  
 \tilde{\tilde V}_4 \!-\! \tilde{\tilde V}_5 + \tilde{\tilde V}_6\right)(x_3)  
 \Bigg]  
 \nn \\  
 && {}+ \frac{M^2 (x_3^0 )^2}{Q^2 + (x_3^0)^2 M^2}
  e^{-(s_0-M^2)/M_B^2}  
 \Bigg[\left(2 \tilde V_1 \!-\! \tilde V_2 \!-\! \tilde V_3 \!-\! \tilde V_4  
 \!-\! \tilde V_5 
 - \frac{2}{x_3}\V_1^{M(d)}
 \right)(x_3^0)  
 \nn \\ && \hspace*{2cm} {}
 + \frac{M^2}{M_B^2}  
 \left(\tilde{\tilde V}_1 \!-\! \tilde{\tilde V}_2 \!-\! \tilde{\tilde V}_3 \!-\!  
 \tilde{\tilde V}_4 \!-\! \tilde{\tilde V}_5 + \tilde{\tilde V}_6\right)(x_3^0)  
  \nn \\ && \hspace*{2cm} {}  
 - M^2 \frac{\dd}{\dd x_3^0}  
 \left(\frac{(x_3^0)^2  
 \left(\tilde{\tilde V}_1 \!-\! \tilde{\tilde V}_2 \!-\! \tilde{\tilde V}_3 \!-\!  
 \tilde{\tilde V}_4 \!-\! \tilde{\tilde V}_5 +  
 \tilde{\tilde V}_6\right)(x_3^0)}{Q^2 + (x_3^0)^2 M^2}\right)  
 \Bigg]  
 \Bigg\}  
 \nn \\ &+&  
 \frac{2 e_u}{f_N}  
 \bigg\{x_3\leftrightarrow x_2,\,\V_1^{M(d)}\to\V_1^{M(u)}\bigg\}\,,  
 \eeqar  
 and
 \beqar{F2}  
 F^{\rm p}_2(Q^2) &=&  
 \frac{e_d}{f_N} \int_{x_3^0}^1 \frac{\dd x_3}{x_3}  
 \exp{  
 \left(- \frac{ \bar x_3 Q^2}{x_3 M_B^2} +  \frac{x_3 M^2}{M_B^2}\right)}  
 \Bigg[\frac{2 M^2}{M_B^2}  
 \left(\tilde V_2 + \tilde V_3 \!-\! \tilde V_1\right)(x_3)  
 \nn \\
 &&{}\hspace*{2cm}
 - \frac{2 M^4}{M_B^4}  
 \left(\tilde{\tilde V}_1 \!-\! \tilde{\tilde V}_2 \!-\! \tilde{\tilde V}_3 \!-\!  
 \tilde{\tilde V}_4 \!-\! \tilde{\tilde V}_5 + \tilde{\tilde V}_6\right)(x_3^0)  
 \Bigg]  
 \nn \\  
 && + \frac{2 M^2 x_3^0}{Q^2 + (x_3^0)^2 M^2}  
 \left[\left(\tilde V_2 + \tilde V_3  
 \!-\! \tilde V_1\right)\!(x_3^0)  
 - \frac{M^2}{M_B^2}  
 \left(\tilde{\tilde V}_1 \!-\! \tilde{\tilde V}_2 \!-\! \tilde{\tilde V}_3 \!-\!  
 \tilde{\tilde V}_4 \!-\! \tilde{\tilde V}_5 + \tilde{\tilde V}_6\right)\!(x_3^0)  
 \right. \nn \\ && \hspace*{2cm} {}  
 \left.  
 + M^2 x_3^0 \frac{\dd}{\dd x_3^0}\!  
 \left(\frac{x_3^0  
 \left(\tilde{\tilde V}_1 \!-\! \tilde{\tilde V}_2 \!-\! \tilde{\tilde V}_3 \!-\!  
 \tilde{\tilde V}_4 \!-\! \tilde{\tilde V}_5 +  
 \tilde{\tilde V}_6\right)\!(x_3^0)}{Q^2 + (x_3^0)^2 M^2}\right)  
 \right]  
 e^{-(s_0-M^2)/M_B^2}     
 \nn \\ &+&  
 \frac{2 e_u}{f_N}  
 \bigg\{ x_3 \leftrightarrow x_2,\,\V_1^{M(d)}\to \V_1^{M(u)}\bigg\}  
 \,.  
 \eeqar  
 The sum rules in \Gl{F1} and \Gl{F2} present the main result of this paper.

In the limit $Q^2 \to \infty$ the sum rules can be simplified. Employing 
asymptotic wave functions we get the asymptotic behavior
\beqar{htexp}
F_1^{\rm p}(Q^2) &=&  \left[
\left(\frac{37}{3} + 2 \frac{\lambda_1}{f_N}\right)e_d +
\left(\frac{37}{3} - 2 \frac{\lambda_1}{f_N}\right)e_u 
\right] \frac{M^2}{M_B^2}
\nn \\ 
&& \qquad \quad {}\times
\frac{1}{Q^6} \left(
s_0^2 M^2 e^{-(s_0 - M^2)/{M_B^2}}  +
\int_0^{s_0} \dd s  \, s^2 e^{-(s - M^2)/M_B^2}  
\right),  
\nn \\ 
F_2^{\rm p}(Q^2) &=&  -2 \left[
    \left(
    \frac{1}{3} \left( 10 - 11 \frac{M^2}{M_B^2} \right)  
- 2 \frac{\lambda_1}{f_N}\left( 1 + \frac{4}{15} \frac{M^2}{M_B^2} \right)
    \right) e_d 
\right.
\nn \\
&&
\left.
+ 2
     \left(
     5 - \frac{8}{9} \frac{M^2}{M_B^2} 
  + \frac{\lambda_1}{f_N} \left( 1 + \frac{2}{15}\frac{M^2}{M_B^2} \right)  
     \right)e_u 
 \right] \frac{M^2}{M_B^2}
 \nn \\ 
&& \qquad \quad {}\times
\frac{1}{Q^8} \left(
s_0^3 M^2 e^{-(s_0 - M^2)/{M_B^2}}  +
\int_0^{s_0} \dd s  \, s^3 e^{-(s - M^2)/{M_B^2}}  
\right).  
\eeqar  
Note that the contribution $\sim 1/Q^6$ to $F_1(Q^2)$ 
arises (cf. \Gl{ltexpand}), which has two sources:
The constants  $\sim 37/3$ originate from the nucleon mass 
corrections and the terms $\sim \lambda_1/f_N$
correspond to the contributions of higher-twist three-quark operators.
Numerically $ \lambda_1/f_N \sim -5$, see the next Section, so that 
both contributions appear to be of the same order.
To avoid misunderstanding note that to leading order in $\alpha_s$ we are 
dealing with the soft, or Feynman contribution to the form factors only. 
The true asymptotic behavior $F_1\sim 1/Q^4$ can be reproduced in the 
LCSR approach as a part of the $O(\alpha_s^2)$ correction,
a calculation which goes far beyond the tasks of this work, 
see Ref.~\cite{BKM00} for the detailed discussion for the pion.

 \section{Numerical Results and Discussion}  
 \setcounter{equation}{0}  

 For the numerical evaluation of the form factors we used the standard 
 value of the continuum threshold $s_0=2.25$~GeV$^2$ and varied the 
 Borel parameter in the range $1.5 < M^2_B < 2.5$~GeV$^2$. The results
 appear to be rather stable so that in the Figures given below we take 
 $M_B^2=2$~GeV$^2$ as our standard choice.  
 \begin{figure}[ht]
 \centerline{  \epsfig{file=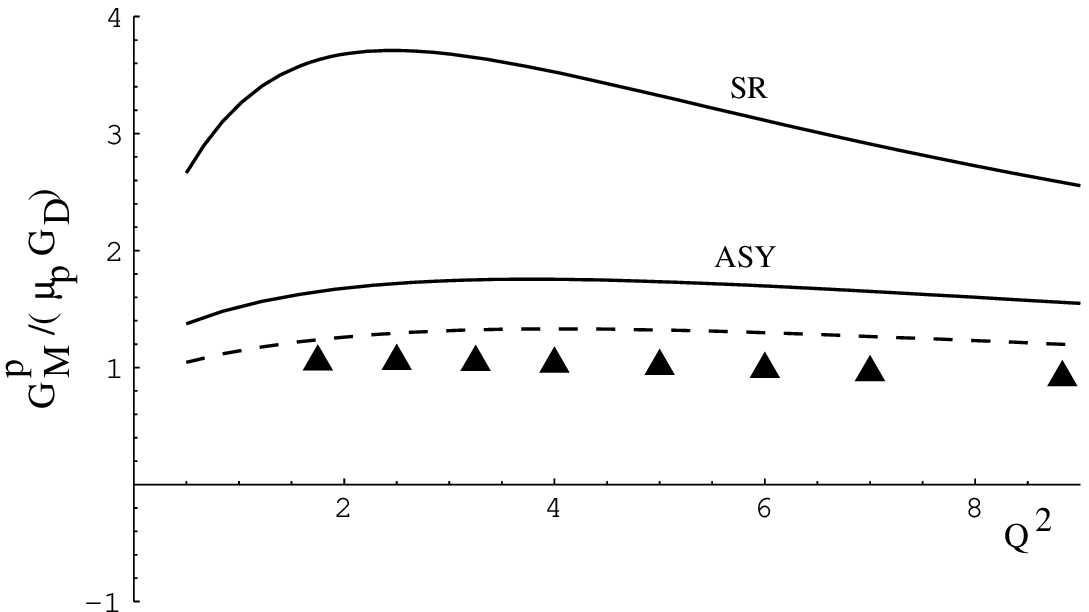,width=7.5cm}\qquad
               \epsfig{file=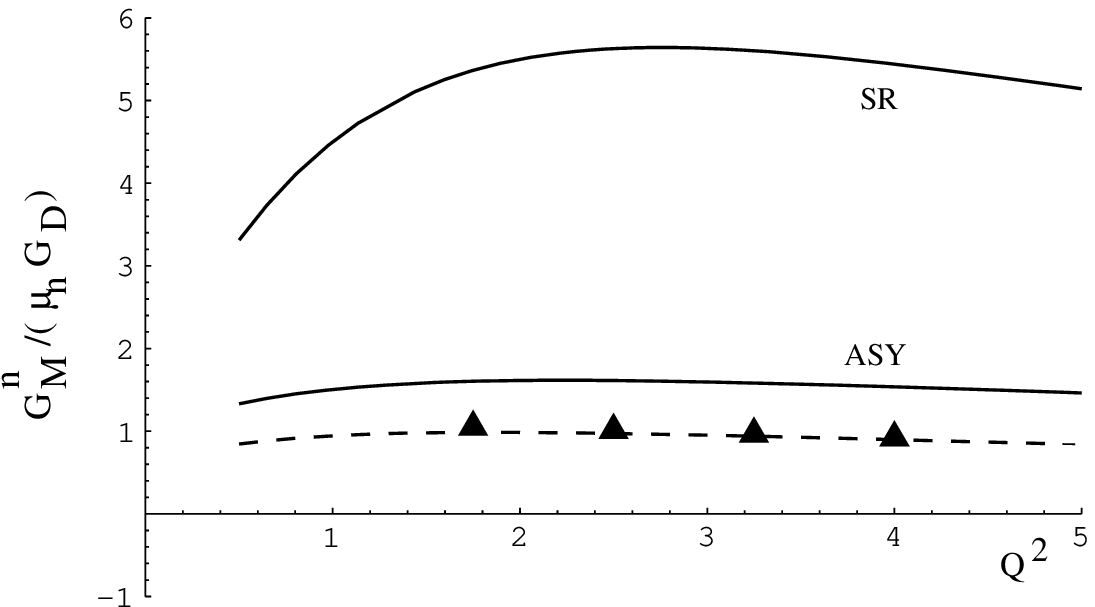,width=7.5cm}  }
 \caption[]{\small\sf 
 The LCSR prediction for  the soft contribution 
 to the magnetic form factor of the
 proton (left)  and the neutron (right).  
 Plotted is the ratio 
 $G_M^{\rm p/n}/(\mu_{\rm p/n} G_D)$ where $G_D$ is the dipole fit 
 \Gl{dipole}. The  solid curves marked {\em `ASY'}\, are obtained 
 using the set of asymptotic distribution amplitudes corresponding to
 the contributions of operators with the lowest conformal spin.
 The curves marked {\em SR} show the results including the next-to-leading
 terms in the conformal expansion with parameters estimated using 
 QCD sum rules. The dashed curves are obtained 
 using asymptotic distribution amplitudes and a reduced relative  
 normalization of the higher-twist contributions within the error range,
 see text.
 The data points are taken from \cite{protonGM,neutron}.}  
 \label{figure3}  

 \vspace{1cm}

 \centerline{  \epsfig{file=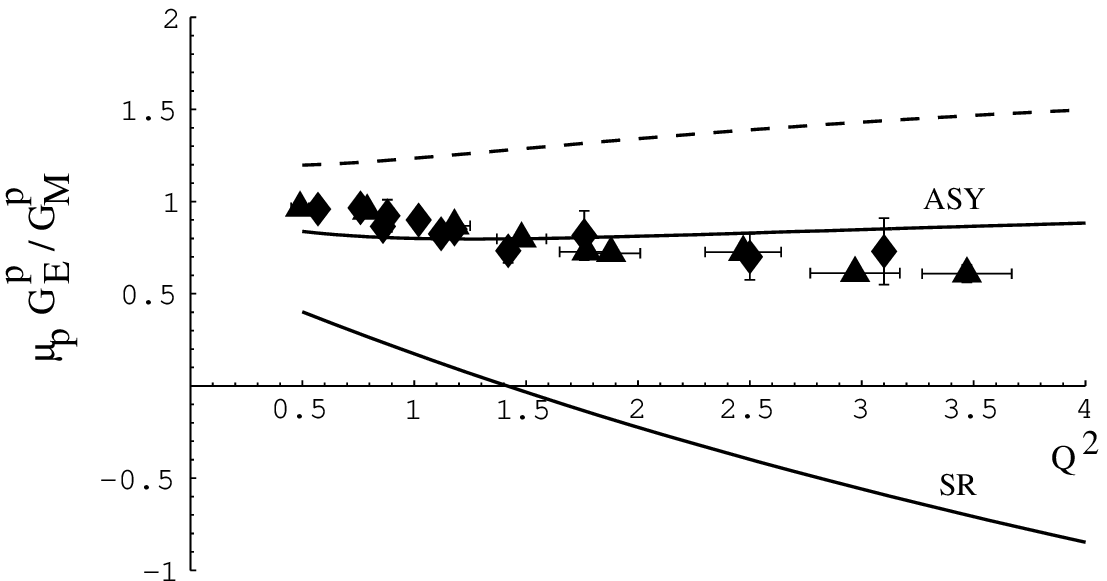,width=7.5cm}\qquad
               \epsfig{file=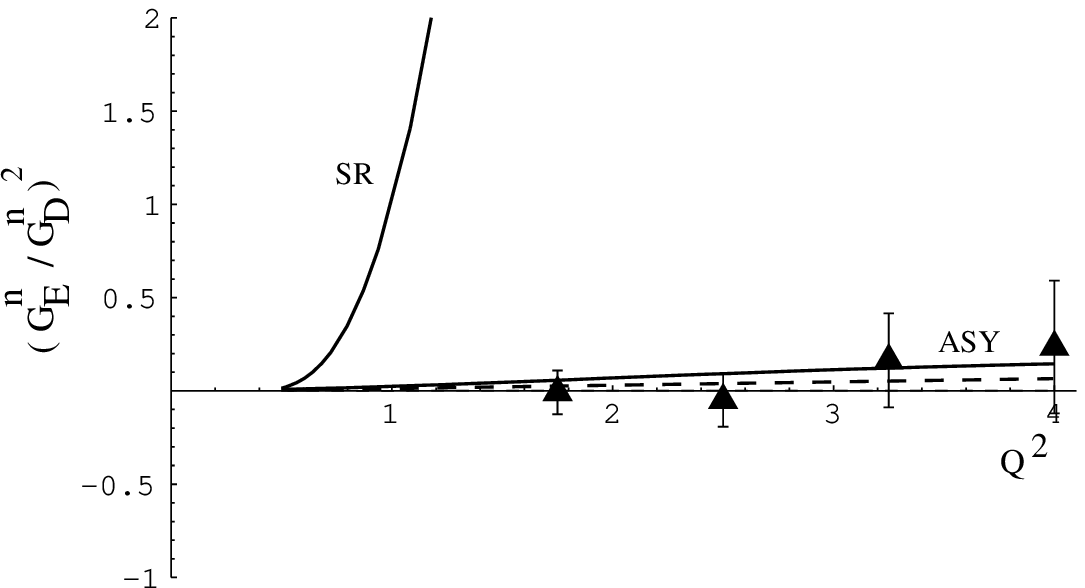,width=7.5cm}  }
 \caption[]{\small\sf 
 The LCSR prediction for  the electric form factor.  
 Plotted is the ratio $\mu_p G_E^{\rm p}/G_M^{\rm p}$ 
 for the proton (left) and $(G_E^{\rm n}/G_D)^2$ for the 
 neutron (right), where  $G_D$ is the dipole formula \Gl{dipole}.  
 The identification of the curves is the same as in \Ab{figure3}.
 The data points are taken from \cite{JLab00,GEproton,neutron}.}  
 \label{figure4}  
 \end{figure}
 The  solid curves marked {\em `asy'}\, in \Ab{figure3} and 
 \Ab{figure4} are obtained using the set of asymptotic distribution 
 amplitudes corresponding  to the contributions of operators with the 
 lowest conformal spin. As already discussed in Sec.~2, the dependence 
 on the normalization $f_N$ of the leading-twist amplitude actually 
 drops out of the sum rules so that to this accuracy the form factors 
 depend on a single nonperturbative parameter - the ratio 
 of matrix elements of the twist-4 and twist-3 operators 
 \beqar{lam/fn}
       \frac{\lambda_1}{f_N} = -5.1 \pm 1.7\,.
 \eeqar 
 The constant $f_N$ is defined in \Gl{local-operators} while $\lambda_1$
 is the familiar nucleon coupling to the so-called Ioffe current \cite{Ioffe81}:
 \beqar{lam1}  
 \eta_{\rm I}(0) &=& \ep^{ijk} \left[u^i(0) C \gamma_\mu u^j(0)\right] \,  
 \ga_5 \gamma^\mu d^k(0) \,, 
 \nonumber\\ 
  \bra{0} \eta_{\rm I}  \ket{P} & = & \lambda_1\, M\, N(P)\,.  
 \eeqar  
 The number in \Gl{lam/fn} is the QCD sum rule estimate
 at the scale 1 GeV (see \cite{BFMS} and references therein).
 Notice that the Ioffe coupling $\lambda_1$ is large compared to 
 $f_N$. Hence the contribution of three-quark states with a 
 different helicity structure compared to leading twist is numerically 
 most important for the form factors in a few GeV range.  

 Beyond the leading order in the conformal expansion the parameters 
 become more numerous and less known. The curves in \Ab{figure3} and 
 \Ab{figure4} marked {\em `SR'}\, are obtained including the contributions
 of the next-to-the leading spin (``P-wave'') and using QCD sum rule 
 estimates \cite{Che84,BFMS} for the four additional parameters that 
 enter the sum rules to this accuracy. For the leading twist this 
 approximation corresponds to the expression in the second line in 
 \Gl{da-twist-3},  and $\tilde\phi^+_3$ in \Gl{twist3par} is the first of
 the additional parameters. Among the remaining three 
 parameters one is of twist-3 and two are of twist-4, see Appendix A for 
 the details.
 It is seen that with large ``P-wave'' contributions 
 the description of the form factors becomes much worse compared to the 
 asymptotic distributions. 
 This is not unexpected, since from the experience
 of calculations of the pion form factor it is known that QCD sum rules
 based on the local operator product expansion tend to overestimate the 
 higher spin corrections considerably, see e.g. \cite{nonlocal}.
 On this evidence, we conclude that   
 large corrections to the asymptotic distribution amplitudes 
 of higher twists are unlikely. 

 The agreement with the experimental data generally becomes better
 if the ratio $\lambda_1/f_N$ is chosen to be 30\% lower compared 
 to the estimate in \Gl{lam/fn}, which corresponds to
   the bottom of the given error range.  
 The corresponding results are shown in in \Ab{figure3} and 
 \Ab{figure4} by the dashed curves. At the same time, we have checked that 
 the agreement cannot be improved substantially by adding 
 ``P-wave'' contributions to the distribution amplitudes with their 
 normalizations taken as free parameters. This, again, can be considered
 as an argument against large corrections to the asymptotic distributions.
  On the other hand, we believe 
 that quantitative conclusions can only be made after the calculation 
 of radiative $O(\alpha_s)$ corrections to the sum rules, which goes 
 beyond the tasks of this paper.    

 To summarize, in this work we have set up a framework for the calculation of 
 baryon electromagnetic form factors in the light-cone sum rule approach. 
 The light-cone  sum rules  
 are derived to leading order in the strong coupling 
 and confronted  with the experimental data.
 We argue that soft non-factorisable terms give a significant contribution  
 to the baryon form factors at intermediate  momentum transfers. 
 Among them, contributions of three-quark states with different  
 helicity structure compared to the leading twist prove to be  
 the most important. A quantitative analysis of baryon form factors 
 requires the calculation of radiative corrections to the sum rule, which is 
 a challenging but doable task. The approach can easily be generalized to 
 the study of transition form factors like $\Delta \to N\gamma$ and 
 to weak decays of heavy baryons.

\appendix  
\renewcommand{\theequation}{\Alph{section}.\arabic{equation}}  
\setcounter{table}{0}  
\renewcommand{\thetable}{\Alph{table}}  
   
\appendix  
\section*{Appendices}  

\section{Summary of Nucleon Distribution Amplitudes}  
\label{app:a}  
\setcounter{equation}{0}  
\setcounter{table}{0}  
  
To make the paper self-contained we collect  here the 
necessary information on the set of nucleon distribution amplitudes 
that enter the LCSRs in \Gl{F1} and \Gl{F1} for the nucleon form factors. 
For definiteness, we consider the proton distribution amplitudes.
The presentation in this Appendix follows Ref.~\cite{BFMS}.

The nucleon distribution amplitudes 
can be classified according to twist in the light-cone quantization approach
of \cite{Kog70}. Hereby the quark field operators  
 are decomposed in ``good'' and ``bad'' or ``plus'' and
``minus'' components, respectively: $q = q^+ + q^-$. 
To this end it is useful to introduce two light-like vectors 
$p = p_+ ,z = z_-$, $p^2=z^2=0$, where $p_\mu = P_\mu + {\cal O}(M^2)$.
The projection operators on the ``plus'' and ``minus'' 
components are $\Lambda_+ = \not\!p\!\not\!z/(2pz)$ and
$\Lambda_- = \not\!z\!\not\!p/(2pz)$, respectively. 
The leading twist-3 amplitude is identified as the one containing three
``plus'' quark fields while each ``minus'' component introduces an
additional unit of twist. This classification is adapted to the applications
in hard reactions: Higher twist amplitudes in general produce 
power-suppressed corrections to physical cross sections.

The distribution amplitudes $V_1(x_i), \ldots, V_6(x_i)$  in~\Gl{opev} 
can be defined \cite{earlybaryon,Che84,BFMS} as nucleon-to-vacuum transition
matrix elements of non-local operators  built of quark fields with definite 
helicity 
\beqar{chiral}
q^{\up(\down)} = \frac{1}{2}(1 \pm \ga_5) q
\eeqar
and separated by light-like distances. In particular, the leading twist-3 
distribution amplitude $V_1$ corresponding to the $(u^+ u^+ d^+)$ component
of the three-quark operator can be written as 
\beqar{vector-twist-3}
\hspace*{-1cm}
\bra{0} \ep^{ijk}\!
\bigg[u_i(a_1 z) C \!\!\not\!{z} u_j(a_2 z)\bigg]_{\ket{10}}
\!\not\!{z} d^{\up}_k(a_3 z) \ket{P}
&=& - \frac12 pz \!\not\!{z} N^\up\! \!\int\!\! \D x
\,e^{-i pz \sum x_i a_i}\,
V_1(x_i)\,.
\eeqar
Here and below we use the notation
\beqar{diquark}
\bigg[u_i(a_1 z) C \!\!\not\!{z} u_j(a_2 z)\bigg]_{\ket{10}}
= \frac12 \left(
u_i^{\up}(a_1 z) C \!\!\not\!{z} u_j^{\down}(a_2 z)
+ u_i^{\down}(a_1 z) C \!\!\not\!{z} u_j^{\up}(a_2 z)
\right)\, .
\eeqar
There exist two twist-4 distributions:
\beqar{vector-twist-4}
\hspace*{-0.3cm}\bra{0} \ep^{ijk}\!
\bigg[u_i(a_1 z) C \!\!\not\!{z} u_j(a_2 z)\bigg]_{\ket{10}}
\!\not\!{p} d^{\up}_k(a_3 z) \ket{P}
&=& - \frac12 pz \!\not\!{p} N^\up\!\! \int\! \D x
\,e^{-i pz \sum x_i a_i}\,
V_2(x_i)\,,
\nn \\
\hspace*{-0.3cm}\bra{0} \ep^{ijk}\!
\bigg[u_i(a_1 z) C\!\not\!{z}\ga_\perp\!\!\not\!{p}
 u_j(a_2 z)\bigg]_{\ket{10}}
\ga^\perp\! \!\!\not\!{z} d^{\down}_k(a_3 z) \ket{P}
&=&   - pz M \!\!\not\!{z} N^\up\! \int\!\! \D x
\,e^{-i pz \sum x_i a_i}\,
V_3(x_i)\,,
 \nn\\
\eeqar
corresponding to the $(u^+u^+ d^-)$ and $(u^+u^- d^+)$ projections,
respectively. Here $\perp$ stands for the projection
transverse to $z$ and $p$, e.g.
$\gamma_\perp \gamma^\perp = \gamma^\mu g_{\mu\nu}^\perp \gamma^\nu$ with
$g_{\mu\nu}^\perp = g_{\mu\nu}-(p_\mu z_\nu+z_\mu p_\nu)/pz$.
Similarly, the two possible projections with two ``minus'' 
fields:
$(u^-u^- d^+)$ and $(u^-u^+ d^-)$ give rise to two 
distribution amplitudes of twist-5 
\beqar{vector-twist-5}
\hspace*{-0.3cm}\bra{0} \ep^{ijk}\!
\bigg[u_i(a_1 z) C \!\not\!{p} u_j(a_2 z)\bigg]_{\ket{10}}
\!\not\!{z} d^{\up}_k(a_3 z) \ket{P}
&=& - \frac14 M^2 \!\not\!{z} N^\up\! \int\!\! \D x
\,e^{-i pz \sum x_i a_i}\,
V_5(x_i)\,,
\nn \\
\hspace*{-0.3cm}\bra{0} \ep^{ijk}\!
\bigg[u^{}_i(a_1 z) C \!\!\not\!{p}\ga_\perp\!\!\not\!{z}
 u^{}_j(a_2 z)\bigg]_{\ket{10}}
\ga^\perp \!\!\not\!{p}\,
d^{\down}_k(a_3 z) \ket{P}
&=&   -  pz M  \!\not\!{p}\, N^\up \int \D x
\,e^{-i pz \sum x_i a_i}\,
V_4(x_i)\,,
\nn \\
\eeqar
respectively. 
Finally, there exists a single twist-6 three-quark 
distribution amplitude
\beqar{vector-twist-6}
\hspace*{-0.3cm}\bra{0} \ep^{ijk}\!
\bigg[u^{}_i(a_1 z) C \!\!\not\!{p} u^{}_j(a_2 z)\bigg]_{\ket{10}}
\!\not\!{p} d^{\up}_k(a_3 z) \ket{P}
&=& - \frac14 M^2 \!\not\!{p} N^\up\! \int\!\! \D x
\,e^{-i pz \sum x_i a_i}\,
V_6(x_i)\,
\eeqar
corresponding to the $(u^- u^- d^-)$ projection.

The distribution amplitudes are scale-dependent and can be expanded 
in contributions of conformal operators that do not mix with each other 
under the renormalization (to one-loop accuracy). To the next-to-leading 
conformal spin accuracy the expansion reads \cite{BFMS}:
\beqar{conf-expand}  
V_1(x_i,\mu) &=& 120 x_1 x_2 x_3 \left[\phi_3^0(\mu) +  
\phi_3^+(\mu) (1- 3 x_3)\right],  
\nn \\  
V_2(x_i,\mu)  &=& 24 x_1 x_2  
\left[\phi_4^0(\mu)  + \phi_4^+(\mu)  (1- 5 x_3)\right],  
\nn\\  
V_3(x_i,\mu)  &=&  
12 x_3 \left[  
\psi_4^0(\mu)(1-x_3)  + \psi_4^-(\mu) (x_1^2 + x_2^2 - x_3  
(1-x_3) ) + \psi_4^+(\mu)( 1\!-\!x_3 \!-\! 10 x_1 x_2)\right],  
\nn \\  
V_4(x_i,\mu) &=& 3 \left[  
\psi_5^0(\mu)(1-x_3)  + \psi_5^-(\mu)\left(2 x_1x_2 - x_3(1-x_3)\right)  
+ \psi_5^+(\mu)(1-x_3 - 2 (x_1^2 +  x_2^2))\right],  
\nn\\  
V_5(x_i,\mu) &=& 6 x_3  
\left[\phi_5^0(\mu)  + \phi_5^+(\mu)(1- 2 x_3)\right],  
\nn\\  
V_6(x_i,\mu) &=& 2 \left[\phi_6^0(\mu) +  \phi_6^+(\mu) (1- 3 x_3)\right],  
\eeqar  
so that it involves 14 parameters. Not all of them, however, are independent: 
They are related to each other by exact QCD equations of motion.

The corresponding analysis was carried out in \cite{BFMS}. It turns out that 
the conformal expansion coefficients in \Gl{conf-expand} can all be expressed 
in terms of 6 independent matrix elements of local operators. To the 
leading conformal spin accuracy (``S-wave'') only two parameters enter, 
$f_N$ and $\lambda_1$, which were defined in \Gl{local-operators} and 
\Gl{lam1}, respectively. One obtains \cite{BFMS}
\beqar{S-wave}  
&& \phi_3^0 = \phi_6^0 = f_N \,,\hspace{0.3cm}  
\qquad  
\phi_4^0 = \phi_5^0 =  
\frac{1}{2} \left(\la_1 + f_N\right) \,,  \hspace{0.3cm} \qquad
\psi_4^0  = \psi_5^0 =  
\frac12\left(f_N - \la_1 \right)  \,.  
\eeqar        
To the ``P-wave'' accuracy four new parameters enter, corresponding 
to matrix elements of the local operators similar to \Gl{local-operators}
($V_1^d, A_1^u$) and \Gl{lam1} ($f_1^d, f_1^u$) but with an additional 
covariant derivative.
The parameters $V_1^d, A_1^u$ are leading twist-3 and, in particular,
the coefficient $\phi_3^+(\mu)$ appearing in the first line in
 \Gl{conf-expand} is given by  
\beqar{3}
 \tilde\phi_3^+ \equiv \phi_3^+/f_N =(7/2)(1- 3 V_1^d)\,.
\eeqar
The parameter $A_1^u$ describes the component in the leading twist distribution
amplitude that is antisymmetric in $x_1 \leftrightarrow x_2$ 
and does not contribute to the sum rules directly. It does 
produce, however,  a contribution to the symmetric parts of higher twist 
amplitudes.
In addition, there are two new parameters $f_1^d, f_1^u$ that are
genuine twist-4. One obtains for twist-4: 
\beqar{4}  
\phi_4^- &=& \frac{5}{4} \left(\la_1(1- 2 f_1^d -4 f_1^u)  
+ f_N( 2 A_1^u - 1)\right) \,,  
\nn \\  
\phi_4^+ &=& \frac{1}{4} \left( \la_1(3- 10 f_1^d)  
- f_N( 10  V_1^d - 3)\right)\,,  
\nn \\  
\psi_4^- &=& - \frac{5}{4} \left(\la_1(2- 7 f_1^d + f_1^u)  
+ f_N(A_1^u + 3 V_1^d - 2)\right) \,,  
\nn \\  
\psi_4^+ &=& - \frac{1}{4} \left(\la_1 (- 2 + 5 f_1^d + 5 f_1^u)  
+ f_N( 2 + 5 A_1^u - 5 V_1^d)\right)\,,  
\eeqar  
for twist-5:  
\beqar{5}  
\phi_5^- &=& \frac{5}{3} \left(\la_1(f_1^d - f_1^u)  
+ f_N( 2 A_1^u - 1)\right) \,,  
\nn \\  
\phi_5^+ &=& - \frac{5}{6} \left(\la_1 (4 f_1^d - 1)  
+ f_N( 3 + 4   V_1^d)\right)\,,  
\nn \\  
\psi_5^- &=& \frac{5}{3} \left(\la_1 (f_1^d - f_1^u)  
+ f_N( 2 - A_1^u - 3 V_1^d)\right)\,,  
\nn \\  
\psi_5^+ &=& -\frac{5}{6} \left(\lambda_1 (- 1 + 2 f_1^d + 2 f_1^u)  
+ f_N( 5 + 2 A_1^u -2 V_1^d)\right)\,,  
\eeqar  
and for twist-6:  
\beqar{6}  
\phi_6^- &=& \phantom{-}\frac{1}{2} \left(\la_1 (1- 4 f_1^d - 2 f_1^u)  
+ f_N(1 +  4 A_1^u )\right) \,,  
\nn \\  
\phi_6^+ &=& - \frac{1}{2}\left(\la_1  (1 - 2 f_1^d)  
+ f_N ( 4 V_1^d - 1)\right)\,.  
\eeqar  
Note that there is a mismatch between the twist classification 
of the distribution amplitudes that implies counting powers of the 
large momentum $p_+$, and the twist classification of the  
local operator matrix elements: E.g. the parameters of the 
twist-4 distribution amplitudes depend both on the 
the leading twist-3 matrix elements $f_N,V_1^d, A_1^u$ and the 
twist-4 matrix elements $\lambda_1,f_1^d,f_1^u$. This 
``propagation'' of lower-twist matrix elements to higher-twist 
distribution amplitudes is
well known and usually referred to as Wandzura-Wilczek contribution.
Note that the distribution amplitudes  of twist-5 and twist-6 are 
entirely of Wandzura-Wilczek type since there exist no geniune 
twist-5 and twist-6 operators to this order in the conformal expansion.     
QCD sum rule estimates for the twist-3 \cite{Che84} and twist-4 \cite{BFMS}
parameters are given in \Ta{tabelle3} together with their asymptotic 
values in the $Q^2\to\infty$ limit.
%
\begin{table}[t]
\renewcommand{\arraystretch}{1.3}
\begin{center}
\begin{tabular}{|c||ccccc|}
\hline
 &$\la_1/f_N$&$V_1^d$&$A_1^u$&$f_1^d$&$f_1^u$
\\
\hline
QCDSR  &$-5.1 \pm 1.7$ & $0.23\pm 0.03$&$0.38\pm 0.15$&$ 0.6\pm 0.2$ & $0.22\pm 0.15 $\\
asymptotic&$-5.1 \pm 1.7$ &   1/3 &  0 & 3/10   & 1/10   \\
\hline
\end{tabular}  
\end{center}
\caption[]{\sf Numerical values for the expansion parameters of the 
nucleon distribution amplitudes.} 
\label{tabelle3} 
\renewcommand{\arraystretch}{1.0}
\label{koeffizienten}
\end{table}

\section{The $O(x^2)$ Corrections}  
\label{app:b}  
\setcounter{equation}{0}  
  
In this Appendix we present the calculation of the $O(x^2)$ corrections
to the light-cone expansion of the three-quark operator in \Gl{zerl}. 
We consider the vector Lorentz projection as it is the only one 
that contributes to the sum rules. We also make use of the important
simplification that to leading order in the strong coupling only the 
bi-local operator with either 
$d$ or $u$ quark shifted from the zero space-time point enters, 
as can be seen from \Gl{calc1}.  
Consider first the $d$-quark contribution:  
\beqar{dquark}  
x^\al \bra{0} \ep^{ijk} \left[u^i  
C\gamma_\alpha u^j\right](0) d_\ga^{k}(x)
          \ket{P} &=&   - x^\al \Bigg[  
\left({\cal V}_1  + \frac{x^2M^2}{4} \V_1^{M(d)}\right) P_\al \left(\ga_5 N\right)_\ga +  
\V_2 M P_\al  
\left(\!\not\!{x} \ga_5 N\right)_\ga  
\nn \\ &&{}\hspace*{-2cm} +  
\V_3 M  \left(\ga_\al \ga_5 N\right)_\ga  
+ \V_4 M^2   x_\al \left(\ga_5 N\right)_\ga  
+ \V_6 M^3  x_\al \left(\!\not{x} \ga_5 N\right)_\ga  
\Bigg].  
\eeqar  
We remind that $\V_1$ starts at leading twist-3, and hence $\V_1^{M(d)}$
is of twist-5. We do not show the $O(x^2)$ corrections to the other Lorentz
structures since they are yet higher twist and will be omitted in what 
follows%
\footnote{Strictly speaking, since we are not taking into account twist-6
contributions induced by $O(x^2)$ corrections to $\V_2$ and $\V_3$, in order
to be consistent we have to discard the contribution of $\V_6$ altogether.
This contribution to the sum rules appears to be numerically neglible, 
however.}.
The invariant functions $\V_1, \ldots, \V_6$ can easily be related to the 
distribution amplitudes of definite twist by taking the light-cone limit 
of \Gl{dquark}, $x^2\to 0$, see \Gl{opev} \cite{BFMS}.
On the other hand, the calculation of  $\V_1^{M(d)}$ is not that immediate 
since in the light-cone limit this contribution vanishes. 
The meaning of the separation of $\V_1$ and $\V_1^{M(d)}$ is most easily 
understood upon the short distance expansion $x_\mu \to 0$. In this 
way, the nonlocal ``string'' operator in the l.h.s. of \Gl{dquark} is 
Taylor-expanded in a series of local operators with three quark fields 
and the increasing number of (covariant) derivatives acting on the d-quark.
The separation of the leading twist part of each local operator 
corresponds to the symmetrisation over all Lorentz indices and  
the subtraction of traces. Without loss of generality, we can consider the 
matrix element contracted with an additional factor $x_\alpha$, 
see \Gl{dquark}, so that the symmetrization is achieved. To subtract
the traces, we formally write 
\beqar{dexpansion}  
\left. x_\al d(x) \right|_{\rm lt}
= \sum_{n= 0}^\infty \frac{1}{n!}  
\Bigg[x_\al x_{\mu_1} \ldots x_{\mu_n} - \frac{x^2}{4}  
\left(\frac{2}{n+1}\right) \sum_{\mu_i,\mu_j}  
\left(x_{\al} \ldots g_{\mu_i \mu_j} \ldots x_{\mu_n}\right)\Bigg]  
\partial^{\al} \ldots \partial^{\mu_n} d(0) \,,
\nn \\  
\eeqar  
where `lt' stands for the leading-twist part.
Observing that $\frac{1}{n+1} = \int_0^1 \dd t \,  t^n $ 
the subtracted contributions $O(x^2)$   
can be reassembled in the form of a non-local string operator:
\beqar{tracesubtract}  
\bra{0} \ep^{ijk} \left[u^i C \!\not\!{x} u^j\right](0) d_\ga^{k}(x)  
          \ket{P}  
&=&  
\bra{0} \left[  
\ep^{ijk} \left[u^i  
C \!\not\!{x} u^j\right](0) d_\ga^{k}(x)  
\right]_{\rm l-t}  
\ket{P}  
\nn \\ &+&  
\frac{x^2}{4} \int_0^1 \dd t  
\frac{\partial^2}{\partial x_\al \partial x^\al}  
\bra{0} \ep^{ijk} \left[u^i  
C \!\not\!{x} u^j\right](0) d_\ga^{k}(t x)  
          \ket{P}.
\eeqar  
The same result can be obtained by observing  \cite{BB89} 
that the leading-twist nonlocal operator has to satisfy   
the homogenous Laplace equation
\beqar{laplace}  
\frac{\partial^2}{\partial x_\la \partial x^\la}  
\bra{0} \left[  
\ep^{ijk} \left[u^i  
C \!\not\!{x} u^j\right](0) d_\ga^{k}(x)  
\right]_{\rm lt}  
\ket{P}  
 = 0 \, .  
\eeqar  
Using QCD equations of motion the second line in \Gl{tracesubtract}  
can be simplified to   
\beqar{eqom1}  
\frac{\partial^2}{\partial x_\al \partial x^\al}  
\ep^{ijk} \left[u^i  
C \!\not\!{x} u^j\right](0) d_\ga^{k}(t x)  
=  
2 t \,  
\ep^{ijk} \left[u^i C \ga^\al u^j\right](0) D_\al d_\ga^{k}(t x)  
+ \mbox{\rm gluons}  
\nn \\  
=  2 t \, \partial_\al  
\ep^{ijk} \left[u^i C \ga^\al u^j\right](0) d_\ga^{k}(t x)  
+ \mbox{\rm gluons}  \, ,
\eeqar  
where $\partial_\alpha$ is a derivative with respect to the overall
translation \cite{BB89};  
for the matrix element we can make the substitution  
$\partial_\al \to - i P_\al$.  
Inserting this result in \Gl{tracesubtract} we finally obtain  
\beqar{tracesubtract2}  
\lefteqn{\hspace*{-1cm}\bra{0} \ep^{ijk} \left[u^i  
C \!\not\!{x} u^j\right](0) d_\ga^{k}(x)  
          \ket{P}  
\,=\,  
\bra{0} \left[  
\ep^{ijk} \left[u^i  
C \!\not\!{x} u^j\right](0) d_\ga^{k}(x)  
\right]_{\rm lt}  
\ket{P} }  
\nn \\ &&{}\hspace*{3cm}  
+ \frac{x^2}{4} (-i 2 P_\al) \int_0^1 \dd t \, t  
\bra{0} \ep^{ijk} \left[u^i  
C \ga^\al u^j\right](0) d_\ga^{k}(t x)  
          \ket{P} + \mbox{\rm gluons} \,.  
\eeqar  
Notice that the r.h.s. only involves (up to corrections with additional 
gluons) the already known distribution amplitudes.  
This equation therefore allows us to determine $\V^{M(d)}$  
--- which appears on the l.h.s. of \Gl{tracesubtract2} ---  
up to gluonic corrections.  

First, consider the leading twist contribution.
We can write  
\beqar{maelements}  
\bra{0} \left[  \ep^{ijk} \left[u^i  
C \!\not\!{x} u^j\right](0) d_\ga^{k}(x)  
\right]_{\rm lt} \ket{P}  
 &=& -   
\int \D x \left[e^{-i P\cdot x x_3} P\cdot x \right]_{\rm lt}  
{\cal V}_1  
\left(\ga_5 N\right)_\ga 
\nn \\
&&{}-  
\int \D x \left[e^{-i P\cdot x x_3} \left(\!\not\!{x} \ga_5 N\right)_\ga  
\right]_{\rm lt}  
P\cdot x \V_2 M  
\nn \\ && -  
\int \D x \left[e^{-i P\cdot x x_3}  \left(\!\not\!{x}\ga_5 N\right)_\ga  
\right]_{\rm lt} \V_3  
M   
+ \ldots  
\eeqar  
where \cite{BB91}  
$\left[e^{-i P\cdot x x_3} (P\cdot x) \right]_{\rm lt}$ and  
$\left[e^{-i P\cdot x x_3} \!\not\!{x}\right]_{\rm lt}$ are   
the leading-twist components for the free fields, defined as the 
 solutions of the corresponding homogeneous Laplace equation.
Note that the factor $P\cdot x$ in the second line in \Gl{maelements}
is not included under the $[\ldots]_{\rm lt}$ bracket since 
$(P\cdot x) \V_2 = 1/2 ( V_1 - V_2 - V_3)$ is a function of 
momentum fractions only and does not contain
any dependence on the position vector $x$. 
The solution can easily be constructed order by order in the 
$(M^2 x^2)^n$ expansion. To the required ${\cal O}(x^2)$ accuracy, the result reads
\cite{BB91}:
\beqar{traceless}  
\left[e^{-i P\cdot x x_3} (Px) \right]_{\rm lt}  
&=& (Px) \left[  
e^{-i P\cdot x x_3} + \frac{x^2 M^2 x_3^2}{4}  
\int_0^1 \dd t \, e^{-i P\cdot x x_3 t}\right]  
\, ,
\\  
\left[e^{-i P\cdot x x_3} \!\not\!{x}\right]_{\rm lt}  
&=&  
\!\not\!{x}\left[  
e^{-i P\cdot x x_3} + \frac{x^2 M^2 x_3^2}{4}  
\int_0^1 \dd t \, t^2  e^{-i P\cdot x x_3 t}\right]  
+  
i \!\not\!{P} \frac{x_3 x^2}{4} \int_0^1 \dd t \, t e^{-i P\cdot x x_3 t}  
\,.  
\nn   
\eeqar  
The corresponding contribution to $\V_1^{M(d)}$ is proportional to the 
nucleon mass squared and involves the leading twist distribution amplitude,
being an exact analogue of the Nachtmann power suppressed correction 
in deep inelastic scattering. The second contribution on the 
r.h.s. in \Gl{tracesubtract2} is special for the exclusive kinematics since
it involves a derivative over the total translation that vanishes for 
forward matrix elements. Its explicit form  
is easily found by contracting the three-quark matrix element  
in \Gl{dquark} with $P_\al$ instead of $x_\al$ and inserting the resulting 
expression  in  \Gl{tracesubtract2}. One gets 
\beqar{translationterm}  
 \lefteqn{ \hspace*{-2cm}\frac{x^2}{4} (-i 2 P_\al) \int_0^1 \dd t \, t  
 \bra{0} \ep^{ijk} \left[u^i  
\ga^\al u^j\right](0) d_\ga^{k}(t x) \ket{P} }
\nn \\    
&& \hspace*{2cm}{}=\frac{x^2 M^2 }{4} i \int {\cal D}x  \int_0^1 \dd t \, t  
e^{- i P\cdot x x_3 t} (V_1 + V_5) (\gamma_5 N)_\gamma + \ldots  
\, ,  
\eeqar  
where the ellipses stand for other Lorentz structures that do not contribute  
to $\V_1^{M(d)}$.  
Inserting everything into \Gl{tracesubtract} we arrive at  
\beqar{endergebnis}  
(Px) \int \dd x_3 e^{-i x_3 P\cdot x)}\V_1^{M(d)}(x_3) &=&  
P\cdot x \int {\cal D} x \, x_3^2 \int_0^1 \dd t \,  
e^{- i P\cdot x x_3 t}\, V_1  
\nn \\ &&  - i  
\int {\cal D} x\,  x_3  \int_0^1 \dd t \, t  
e^{- itx_3 P\cdot x} (V_1 -V_2)  
\nn \\  &&  
+ \frac{1}{Px} \int {\cal D} x  
e^{- i x_3 P\cdot x } (- 2 V_1 + V_3 + V_4 + 2 V_5)  
   \, .  
\eeqar  
In order to solve this equation we expand both sides
at short distances and obtain the moments of  $\V_1^{M(d)}$
with respect to $x_3$ expressed through 
moments of the  distribution amplitudes  
defined as $V_i^{(d)(n)} = \int {\cal D} x \, x_3^n V_i(x_i)$.  
One finds  
\beqar{Vd-moments}  
\int \dd  x_3 \,x_3^n \,\V_1^{M(d)}(x_3)  
&=& 
- \frac{1}{(n+1)(n+2)} \bigg[(- 2 V_1 + V_3 + V_4 + 2 V_5)^{(d)(n+2)}  
\bigg]                 
\\  &&\hspace*{-1.5cm}{} + 
\frac{1}{(n+1)(n+3)} \left[ (n+3) V_1^{(d)(n+2)}  
- \frac12 (V_1 -V_2)^{(d)(n+2)} - (V_1 +V_5)^{(d)(n+1)}\right],
\nn 
\eeqar
up to contributions of multiparton distribution amplitudes with extra
gluons that have been neglected. 
  
The analysis of the $u$-quark contribution is performed in a 
similar way. We consider the matrix element
\beqar{uquark}  
x^\al \bra{0} \ep^{ijk} \left[u^i(0)  
C\gamma_\alpha u^j(x)\right] d_\ga^{k}(0)  
          \ket{P} &=&  - x^\al \Bigg[  
\left({\cal V}_1  + \frac{x^2 M^2}{4} 
\V_1^{M(u)}\right) P_\al \left(\ga_5 N\right)_\ga   
\nn \\ && \hspace*{-7cm}{}+  
\V_2 M P_\al  
\left(\!\not\!{x} \ga_5 N\right)_\ga  +
\V_3 M  \left(\ga_\al \ga_5 N\right)_\ga  
+ \V_4 M^2   x_\al \left(\ga_5 N\right)_\ga  
+ \V_6 M^3  x_\al \left(\!\not{x} \ga_5 N\right)_\ga  
\Bigg]  
\eeqar  
and find repeating the same steps that lead to \Gl{tracesubtract}:  
\beqar{tracesubtractu}  
\bra{0} \ep^{ijk} \left[u^i(0)  
C \!\not\!{x} u^j(x)\right] d_\ga^{k}(0)  
          \ket{P}  
&=&  
\bra{0} \left[  
\ep^{ijk} \left[u^i(0)  
C \!\not\!{x} u^j(x)\right] d_\ga^{k}(0)  
\right]_{\rm lt}  
\ket{P}  
\nn \\ &&  
+ \frac{x^2}{4} \int_0^1 \dd t  
\frac{\partial^2}{\partial x_\al \partial x^\al}  
\bra{0} \ep^{ijk} \left[u^i(0)  
C \!\not\!{x} u^j(t x)\right] d_\ga^{k}(0)  
          \ket{P}  
\nn \\ &=&  
\bra{0} \left[  
\ep^{ijk} \left[u^i(0)  
C \!\not\!{x} u^j(x)\right] d_\ga^{k}(0)  
\right]_{\rm lt}  
\ket{P}  
+ \mbox{\rm gluons},
\eeqar  
the only difference being that the term  
corresponding to a total translation does not arise in this case.   
{}For the  moments with respect to $x_2$ we get  
\beqar{Mu-moments}  
\int \dd x_2  \,x_2^n \,\V_1^{M(u)}(x_2) 
               &=& \frac{1}{(n+1)(n+3)} \left[ (n+3) V_1^{(u)(n+2)}  
- \frac12 (V_1 -V_2)^{(u)(n+2)} \right]  
\nn \\ && 
- \frac{1}{(n+1)(n+2)} \bigg[(- 2 V_1 + V_3 + V_4 + 2 V_5)^{(u)(n+2)}  
\bigg]  
\, .  
\eeqar  
The corresponding expressions in the momentum fraction space are easily 
obtained by inserting the conformal expansions for   
$V_1, \ldots, V_6$ and inverting the moment equations. The result reads
\beqar{VM}
\V_1^{M(u)}(x_2) 
  & = & \frac{x_2^2}{24}\left( \lambda_1 C_{\lambda}^u + f_N C_{f}^u \right)\,,
\\
\V_1^{M(d)}(x_3) 
  & = & \frac{x_3^2}{24}\left( \lambda_1 C_{\lambda}^d + f_N C_{f}^d \right)
\nn
\eeqar
with
\beqar{VM2}
C_{\lambda}^u & = & - (1 - x_2)^3 
              \left[13 - 20f^d_1 + 3x_2 + 10f^u_1(1 - 3x_2)\right]\,, 
\nonumber
\\
C_{f}^u       & = & (1 - x_2)^3 \left[113 + 495x_2 - 552x_2^2 + 
              10A^u_1(-1 + 3x_2) +  2V^d_1(113 - 951x_2 + 828x_2^2) \right]\,, 
\nonumber
\\
C_{\lambda}^d & = &  - (1-x_3) \left[11 + 131x_3 - 169x_3^2 + 63x_3^3 - 
    30f^d_1(3 + 11x_3 - 17x_3^2 + 7x_3^3) \right]  
\nonumber
\\
& & 
- 12(3 - 10f^d_1) \ln[x_3]\,,
\nonumber
\\
C_{f}^d       & = & - (1 - x_3) \left[1441 + 505x_3 - 3371x_3^2 + 3405x_3^3 - 1104x_3^4 \right.
\nonumber
\\
& & \left.
 -  24V^d_1(207 - 3x_3 - 368x_3^2 + 412x_3^3 - 138x_3^4) \right] - 12(73 - 220V^d_1) \ln[x_3]\,.
\eeqar


\end{document}